\documentclass[conference, letterpaper,onecolumn]{IEEEtran}

\usepackage[utf8]{inputenc} 
\usepackage[T1]{fontenc}    
\usepackage{hyperref}       
\usepackage{url}            
\usepackage{booktabs}       
\usepackage{amsfonts}       
\usepackage{nicefrac}       
\usepackage{microtype}      
\usepackage{lipsum}
\usepackage{fancyhdr}       
\usepackage{graphicx}       
\graphicspath{{media/}}     

\usepackage[utf8]{inputenc} 
\usepackage[T1]{fontenc}
\usepackage{url}              
\usepackage{cite}             

\usepackage[cmex10]{amsmath}  
\usepackage{graphicx}         
\usepackage{booktabs}         

\usepackage{float}
\usepackage{caption}
\usepackage[utf8]{inputenc} 
\usepackage[T1]{fontenc}
\usepackage{url}
\usepackage{ifthen}
\usepackage{graphicx}
\usepackage{amsthm}
\usepackage{amsfonts}
\usepackage{cite}
\usepackage{multicol}
\usepackage{color}
\usepackage{amssymb}
\usepackage[cmex10]{amsmath} 
\usepackage{stfloats}
\usepackage{enumitem}
\newlist{steps}{enumerate}{1}
\setlist[steps, 1]{label = Step \arabic*:}
\usepackage{subfigure}

\newtheorem{theorem}{Theorem}[section]

\newtheorem{defn}{Definition}
\newtheorem{thm}{{\cal T}heorem}
\newtheorem{cor}{Corollary}
\newtheorem{prop}{Proposition}
\newtheorem{lem}{Lemma}
\newtheorem{conj}{Conjecture}
\newtheorem{constr}{Construction}
\newtheorem{note}{Note}
\newtheorem{example}{Example}
\newtheorem*{example*}{Example}

\newtheorem*{remark*}{Remark}

\newcommand{\bit}{\begin{itemize}}
	\newcommand{\eit}{\end{itemize}}
\newcommand{\bcor}{\begin{cor}}
	\newcommand{\ecor}{\end{cor}}
\newcommand{\beq}{\begin{equation}}
	\newcommand{\eeq}{\end{equation}}
\newcommand{\beqn}{\begin{equation}}
	\newcommand{\eeqn}{\end{equation}}
\newcommand{\bea}{\begin{eqnarray}}
	\newcommand{\eea}{\end{eqnarray}}
\newcommand{\bean}{\begin{eqnarray*}}
	\newcommand{\eean}{\end{eqnarray*}}
\newcommand{\ben}{\begin{enumerate}}
	\newcommand{\een}{\end{enumerate}}
\newcommand{\bdefn}{\begin{defn}}
	\newcommand{\edefn}{\end{defn}}
\newcommand{\bnote}{\begin{note}}
	\newcommand{\enote}{\end{note}}
\newcommand{\bprop}{\begin{prop}}
	\newcommand{\eprop}{\end{prop}}
\newcommand{\blem}{\begin{lem}}
	\newcommand{\elem}{\end{lem}}
\newcommand{\bthm}{\begin{thm}}
	\newcommand{\ethm}{\end{thm}}
\newcommand{\bconj}{\begin{conj}}
	\newcommand{\econj}{\end{conj}}
\newcommand{\bconstr}{\begin{constr}}
	\newcommand{\econstr}{\end{constr}}
\newcommand{\bpf}{\begin{proof}}
	\newcommand{\epf}{\end{proof}}

\newcommand{\baln}{\begin{align*}}
	\newcommand{\ealn}{\end{align*}}
\newcommand{\bal}{\begin{align}}
	\newcommand{\eal}{\end{align}}

\newcommand{\nng}{\mathcal{G}_{k-1}}
\newcommand{\extg}{\mathcal{H}}
\usepackage{tikz}
\newcommand*\circled[1]{\tikz[baseline=(char.base)]{
		\node[shape=circle,draw,inner sep=1pt] (char) {#1};}}

\IEEEoverridecommandlockouts

\begin{document}

\title{Latency-Optimal File Assignment in Geo-Distributed Storage with Preferential Demands}
\author{
	\IEEEauthorblockN{Srivathsa Acharya, P. Vijay Kumar \ \\}
 	\IEEEauthorblockA{
 			Department of Electrical Communication Engg.,\\ IISc, India \\ \{srivatsa.acharya, pvk1729\}@gmail.com}
	\and 
	\IEEEauthorblockN{Viveck R. Cadambe \ \\}
 	\IEEEauthorblockA{
 			School of Electrical Communication Engg.,\\ Georgia Tech, USA \\ viveck.cr@gmail.com}
}
\maketitle

\begin{abstract}
We consider the problem of data storage in a geographically distributed (or geo-distributed) network of servers (or nodes) where inter-node communication incurs certain round-trip delays. Every node serves a set of users who can request any file in the network. If the requested file is not available at the node, it communicates with other nodes to obtain the file, thus causing the user to experience latency in obtaining the file. 
The files can be placed uncoded, where each node stores exact copies of the files, or in coded fashion, where certain linear combination of files are placed at each node.
We aim to obtain an optimal file placement on the nodes with respect to minimizing the worst-case latency at each node, as well as the system-average latency. The prior literature considered the case of equiprobable file demands at the nodes. In this paper, we investigate the generic case of non-uniform file-demand probabilities at each node. The scheme presented here is optimal within the family of uncoded schemes. It is obtained first by modeling the worst-case latency constraint as a vertex coloring problem, and then converting the system-average latency optimization to a problem of balanced-assignment.
\end{abstract}

\section{Introduction}

Distributed data storage systems are essential components of today's cloud-computing infrastructure. The data centers are often placed across a wide geographic area to enable data access to clients across the globe. Some of the major commercial cloud storage providers such as  Google Cloud \cite{Spanner}, Amazon AWS \cite{aws-geo}, and Microsoft Azure \cite{cosmos-geo} offer support for such geographically distributed (or geo-distributed) data storage. In the area of distributed storage, coding theory has contributed significantly in ensuring efficient fault-tolerance of the stored data, for e.g., through the development of regenerating codes \cite{dimakis2010network, wu2009reducing}, locally repairable codes \cite{gopalan2012locality, tamo2014family} and codes with availability \cite{rawat2016locality} (see \cite{balaji2018erasure}, \cite{vinayaketal2022FnT} for a survey). In this paper, we focus on an application of coding theory where data redundancy across geographically distributed servers is primarily designed to provide low latency data access to users.

Geo-distributed storage networks consist of nodes (data-centers/servers) connected to each other through links having certain round-trip delays. Each node serves a specific set of clients, where a client can request for any data available in the network. One of the desired features of geo-distributed storage systems is to provide wait-free or low-latency access to data. However, if the data requested by a client is not available at a parent node, it has to be fetched from remote nodes, thereby incurring a data-access latency of the inter-node round-trip time (RTT). The RTTs between the nodes can be relatively large (tens to a few hundreds of ms, see AWS ping-measurements\cite{aws-geo}) and hence are a dominant component of user data access latency.
 
Providing wait-free access requires every file to be replicated at every node, which is inefficient in terms of storage utilization, and also unfeasible when the storage requirement is comparable to the total storage capacity of the system. Given that total replication  of files at nodes is not possible, several schemes based on \emph{partial replication} have been proposed in the literature, where each node stores only a subset of the data (see \cite{cadambe},\cite{cadambeArxiv} and references therein), which we refer to  as \emph{uncoded} storage schemes. Instead of storing copies of data files across nodes, one could also store linear combination of files in the nodes, which we refer to as \textit{coded storage}.

Design of uncoded storage schemes taking the data-access latency into account is the main focus of this paper. Here, we study two latency metrics that are relevant to practice. First, we consider the worst-case latency incurred at each node - the maximum RTT required to fetch a file from a node for a given storage scheme. The second metric is the average latency (measured across nodes and files), which determines the average throughput of the system  as per Little's law\cite{cadambeArxiv}.

In computer systems and performance analysis literature, there are several works that aim to optimize data placement in geo-distributed data storage systems by utilizing knowledge of inter-node RTTs \cite{ardekani2014self, spanstore, replication-placement, shankaranarayanan2014performance, Abebe2018ECStoreBT, su2016systematic, ec-geo-placement,zare2022legostore,SoljaninEtAl}.  These works develop optimization frameworks and solutions for uncoded/coded placement based on latency, communication cost, storage budget, and fault-tolerance requirements. However, they do not consider the latency minimization as the primary objective.  There are also works which provide latency analysis based on MDS storage (as in \cite{SoljaninEtAl},\cite{lee_shah_huang_ramachandran_2017}), but MDS codes are not suited well for average latency constraints due to significant inter-node communication in decoding the coded files (see~\cite{acharya_etal_LatencyOptimalUncodedStorage_ISIT2024}). 

Thus, unlike the classical storage codes, the codes must be designed and codeword symbols must be placed on the nodes based on the RTTs to minimize latency. In particular, for a given storage budget and worst-case latency, it is unclear when uncoded strategies obtain optimal average latency. The work in  \cite{acharya_etal_LatencyOptimalUncodedStorage_ISIT2024} addressed this question for the case when the file demands are equally likely at each node,  and provided necessary and sufficient conditions for an uncoded storage to be latency optimal. In reality, the file preferences are not uniform, and moreover can vary across geographically distributed nodes. The current work extends to the case of generic file demand probabilities across the nodes.


The rest of the paper is organized as follows. Section~\ref{sec:sys_model} develops the system-model, provides formal definitions of storage codes and associated latency metrics. Section~\ref{sec:vertex_coloring} deals with the worst-case latency constraint where the uncoded storage solution is modeled as one of vertex-coloring of an associated graph, drawn from the work in ~\cite{acharya_etal_LatencyOptimalUncodedStorage_ISIT2024}. Section~\ref{sec:avg_latency} deals with the system-average latency problem in the presence of preferential file demands. 
In Section~\ref{sec:optimal_file_to_color}, we provide the main contribution of the paper where the problem of finding an optimal uncoded scheme given preferential file demands is solved by  modeling it as a \emph{balanced-assignment} problem. Section~\ref{sec:conc} summarizes the contribution of the paper along with directions for future work.

\section{System Model} \label{sec:sys_model}
A geo-distributed storage network is modeled by the tuple $(n, k,T_{n\times n}, P_{n\times k})$ having $n$ nodes and $(k\le n)$ files to be stored across the network. Each node is assumed to store one file\footnote{The applicability of the system model to the general case of nodes storing several files is explained in Section~\ref{subsec:multi-file node}}. The matrix $T = [\tau_{u,v}]_{n\times n}$ is symmetric, where $\tau_{u,v}$ represents the round-trip time (RTT) between the nodes $u,v$. The model thus considers single-hop communication between the nodes\footnote{Multi-hop communication can also be modeled by assigning the least delay path between the two nodes. The RTTs should therefore satisfy the triangle inequality.}. $P = [p_{v,j}]_{n\times k}$ is the non-negative demand-probability matrix where $p_{v,j}$ represents the probability that a file demand corresponds to the $j^{\text{th}}$ file at node $v$. Let $[k] = \{1,2,\dots,k\}$ denote the set of file indices, and let $\mathcal{V}$ be the set of nodes. Then, \(\sum_{v \in \mathcal{V}}\sum_{j\in[k]} p_{v,j} = 1 \). 

We next introduce the notion of a linear storage code. Let the $k$ files be $\{W_1,W_2,\dots W_k\}$, and the node contents be $\{X_1,X_2,\dots,X_n\}$. Assuming that the files belong to a certain finite field $\mathcal{F}$, we define a linear storage code $\mathcal{C}$ by a rank-$k$ generator matrix $G_{k\times n}$ such that
\[\begin{pmatrix}X_1,X_2,\dots, X_n\end{pmatrix} = \begin{pmatrix}W_1,W_2, \dots, W_k\end{pmatrix}G \text{.}\]
An uncoded storage is a special case of the linear storage code given by $X_{v} = W_{\phi(v)}$ for some function $\phi: \mathcal{V}\to [k]$.

We next look into the process of obtaining any of the $k$ files at each node. For this, each node $v$ maintains a recovery matrix $R^{(v)}_{n\times k}$ such that 
\[\begin{pmatrix}W_1,W_2, \dots, W_k\end{pmatrix} = \begin{pmatrix}X_1,X_2,\dots, X_n\end{pmatrix}R^{(v)} \text{.}\]

\noindent Thus, \(GR^{(v)} = I_{k\times k}\). Let $G^{\dagger}_{n\times k}$ be a right inverse of $G$ such that $GG^{\dagger}= I_{k\times k}$. Then \(G(R^{(v)}-G^{\dagger}) = 0_{k\times k}\). This means that the columns of $(R^{(v)}-G^{\dagger})$ belong to the dual code of code $\mathcal{C}$. Denoting $H_{(n-k) \times n}$ to be a  parity-check matrix of the code $\mathcal{C}$ (equivalently, a generator matrix of the dual code) such that $GH^T = 0$, we thus get
\begin{equation} 
        R^{(v)}  - G^{\dagger}  = H^T \text{} A^{(v)}_{(n-k)\times k}\text{,}
\label{eq:recovery matrix}
\end{equation}        
where $A^{(v)}$ is an arbitrary ${(n-k)\times k}$ matrix, the choice of which impacts the user-latency as described next.

Given a node $v$, denote $j^{\text{th}}$ column of the recovery matrix by $R^{(v)}(:,j)$. Since \(W_j = (X_1,\dots , X_n)R^{(v)}(:,j)\), the node $v$ needs files from nodes $\{s\in \mathcal{V}: R^{(v)}(s,j) \ne 0\}$ to recover $W_j$. Thus, the latency incurred at node $v$ to decode file $W_j$ is
    \begin{equation}\ell_{v,j} = \max_{s\in \mathcal{V}} \{\tau_{s,v}:R^{(v)}(s,j) \ne 0\}\text{.}
    \label{eq:latency_from_recovery_matrix}
    \end{equation}
From \eqref{eq:recovery matrix}, \(R^{(v)}(:,j) = G^{\dagger}(:,j) +H^T A^{(v)}(:,j)\). Thus, the latency $\ell_{v,j}$ is in turn determined by $A^{(v)}(:,j)$. The columns of $A^{(v)}$ are chosen to obtain the least latency $\ell_{v,j}$ for each file $j\in [k]$. Henceforth, it is assumed that for a given code $\mathcal{C}$, the optimal $R^{(v)}$ is used at each node $v$ as described above.

We further define below the latency metrics with which we assess the performance of a linear storage code. 
Per-node worst-case latency (WC-latency) is defined as the maximum latency at a node $v$ to decode any of the $k$ files, given by \[\ell_{\max}(v) = \max_{j \in [k]} \ell_{v,j} \text{.}\]
The system-average latency (or average latency) is defined as the expected value of the latencies across all nodes and files: 
\begin{equation}
L_{\text{avg}} = \sum_{v\in\mathcal{V}}\sum_{j \in [k]} \ell_{v,j}p_{v,j}\text{.}
\label{eq:avg_latency}
\end{equation} 
The objective is to find a storage code $\mathcal{C}$ that is optimal with respect to WC-latency $\ell_{\max}(v)$ at each node $v$, as well as the system-average latency $L_{\text{avg}}$. In this context, when we say that a code is optimal, we refer to optimality with respect to both WC-latency  and the average latency.

In some cases (see examples of \cite{acharya_etal_LatencyOptimalUncodedStorage_ISIT2024}), an uncoded storage itself can achieve this objective, while in others, coded storage can outperform any uncoded scheme with respect to minimizing both WC-latency and average latency. Thus, it is important to know the conditions on the geo-distributed storage network for which there exists a latency-optimal uncoded storage.
In \cite{acharya_etal_LatencyOptimalUncodedStorage_ISIT2024}, this was addressed for the case when the file requests are equally likely, that is, when $p_{v,j} = 1/kn$ for all files at all nodes. In this setting, \cite{acharya_etal_LatencyOptimalUncodedStorage_ISIT2024} provided a vertex-coloring based condition that is necessary and sufficient for an optimal uncoded storage to exist. 
In the current work, the focus is instead on optimal uncoded schemes for generic file-demand probabilities given by the matrix $P$. Here, it turns out that the vertex coloring condition for uniform file-demands is necessary, but not sufficient. In addition, the color-to-file assignment impacts the average latency and hence needs to be optimized. This is detailed in the coming sections.
 
\subsection{Multi-File Node Case} \label{subsec:multi-file node}
In the system model, we assumed that nodes store one file each. In general, each node $v$ stores $M_v\ge 1 $ files, and the objective is to store $K$ files where $K\le \sum_{v \in \mathcal{V}}M_v$, the overall storage capacity. This can be converted to an equivalent network of $N = \sum_{v\in\mathcal{V}}M_v$ nodes, with  each node storing one file as follows. 
\begin{enumerate}
    \item Split each node $v$ into $M_v$ sub-nodes each having unit-file capacity.
    \item Connect the sub-nodes of each node pair-wise with $0$-RTT edges.
    \item Connect each sub-node of node $v$ with each sub-node of another node $w$ with an edge with RTT $\tau_{v,w}$. Thus we now have a complete graph of $N = \sum_{v\in\mathcal{V}}M_v$ sub-nodes, where any two sub-nodes are connected by an edge.
    \item To assign the demand probabilities, we assign each sub-node $i$ of a node $v$ with probability 
    \( q_{v_i,j} = \frac{p_{v,j}}{M_v}   \text{ for } i\in [M_v], j\in [K]\).
\end{enumerate}

Thus, we have obtained a $(N,K, Q_{N\times K} = [q_{v_i,j}])$ geo-distributed storage network of sub-nodes with each sub-node storing one file each.  
The equivalence of this network to the original network of $n$ nodes can be seen from the following.
\begin{itemize}
    \item Suppose file assignment is done on each of the sub-nodes, $\{v_i, i \in [M_v]\}$ of node $v$ based on the optimal scheme discussed in Section~\ref{sec:optimal_file_to_color}. Then, on the original network, a node $v$ will be assigned $M_v$ files corresponding to the files assigned to its $M_v$ sub-nodes.
    \item At a node $v$, the RTT to fetch any of its $M_v$ files is $\tau_{v,v}= 0$. Alss, the RTT to fetch any of the $M_w$ files of another node $w$ is $\tau_{v,w}$, respecting the original RTT values. 
    \item The demand probability of a file $W_j$ at node $v$ is $\sum_{i \in [M_v]}q_{v_i,j} = p_{v,j}$, same as the demand probability in the original network. 
\end{itemize}
Hence, it suffices to deal only with unit-file capacity nodes.

\section{WC-Latency Optimality and Vertex Coloring} \label{sec:vertex_coloring}
To handle both WC-latency and average-latency objectives, we first place a constraint on the codes to achieve WC-latency, and then look within only these codes for the least average latency. The WC-latency constraint and resulting vertex-coloring condition is taken from \cite{acharya_etal_LatencyOptimalUncodedStorage_ISIT2024}, and is described next.
Define \[\big( \lambda_{v,0} \le \lambda_{v,1} \le \dots \le \lambda_{v, (n-1)} \big)\] as the sequence of RTTs to node $v$ (that is, $\tau_{u,v}$) in ascending order, with possible repetitions. Note that $\lambda_{v,0} := \tau_{v,v}=0$.
\bprop \label{prop:wc_latency_bound}
For any code $\mathcal{C}$, WC-latency at a node $v$ is bounded as
 \[\ell_{\max}(v) \ge \lambda_{v,(k-1)} \text{.}\]
\eprop
This comes from the fact that a node must contact at least $(k-1)$ nodes to get $k$ files, and hence WC-latency cannot be smaller than the RTT to the $(k-1)^{\text{th}}$ least-RTT neighbor. 

For any geo-distrubted storage network, there exist codes that achieve WC-latency bound, such as the MDS codes described in \cite{acharya_etal_LatencyOptimalUncodedStorage_ISIT2024}. Given this, we focus only on the family of codes that are WC-latency optimal.

Given the recovery matrices $R^{(v)}$ for a certain code $\mathcal{C}$, we define the \emph{recovery graph} as the graph on the node set $\mathcal{V}$ where a directed edge is assigned to each link where file-transfer happens during recovery. That is, for $s\ne v$, $(s,v)$ is a directed edge in the recovery graph whenever \(R^{(v)}(s,j) \ne 0\) for some $j\in [k]$. 
 
\bdefn
The Nearest Neigbhor Graph $\nng$ is a directed graph on the node set $\mathcal{V}$ formed by placing an incoming edge to each node $v$ from its $(k-1)$ least-RTT neighbors.
\edefn

Here, when we say an incoming edge to node $v$ from a node $s \ne v$, we refer to the edge $(s,v)$. Also, we refer to \emph{nearest} in terms of RTT between nodes instead of the distance. For simplicity, we will assume that the RTTs to a node are distinct so that $\nng$ is unique\footnote{When RTTs to a node are not distinct, we may have multiple choices of $(k-1)$ nearest neighbors, and hence multiple $\nng$. This case can also be handled by applying the algorithm in Section~\ref{sec:optimal_file_to_color} on each $\nng$, and then selecting that $\nng$ with the smallest $L_{\text{avg}}$ cost.}. 

\bdefn[Code Admissibility on $\nng$]
A code is \textbf{admissible} on $\mathcal{G}_{k-1}$ if its recovery graph is $\mathcal{G}_{k-1}$.  
\edefn
In other words, a code is admissible on $\nng$ if each node accesses files only from its $(k-1)$ nearest neighbors for obtaining any of the $k$ files. 

\bprop \label{prop:wc_optimality} 
A code is WC-Latency optimal if it is admissible on $\mathcal{G}_{k-1}$. 
\eprop
This is due to the fact that a code admissible on $\nng$ accesses files only from its $(k-1)$ nearest neighbors and hence meets the worst-case latency bound given in Proposition~\ref{prop:wc_latency_bound}.

Therefore, we look for uncoded storage admissible on $\nng$ as they are WC-latency optimal. Related to $\nng$ we define a second graph on node set $\mathcal{V}$ called the extended graph as follows.
\bdefn
An extended graph $\extg$ is an undirected graph on the node set $\mathcal{V}$ obtained by the following steps:
\ben
	\item Connect each node to its $(k-1)$ nearest neighbors.
	\item Further, connect these $(k-1)$ neighbors pair-wise.
\een
\edefn 

A vertex coloring of a graph is an assignment of colors to vertices such that the adjacent nodes (nodes connected by an edge) have different colors \cite{diestel}. If the vertex coloring can be accomplished with $k$-colors, the graph is said to be $k$-colorable.
 
\begin{theorem}[Vertex Coloring Condition]\label{thm:vc}
An uncoded storage scheme is admissible on $\nng$ if and only if the extended graph $\mathcal{H}$ is $k$-colorable.
\end{theorem}
The proof comes by noting that an uncoded storage is admissible on $\nng$ if and only if each node and its $(k-1)$ nearest neighbors have different files. Thus in $\extg$, the adjacent nodes have different files. This is equivalent to $\extg$ being $k$-colorable, by viewing $k$ files as synonymous with $k$ colors.

Thus, Theorem~\ref{thm:vc} along with Proposition~\ref{prop:wc_optimality} provides the existence condition for WC-latency optimal uncoded storage. 

\section{Average Latency Optimality}\label{sec:avg_latency}
\begin{figure}
    \centering
    \subfigure[Uncoded Scheme]{
    \includegraphics[width=0.4\linewidth]{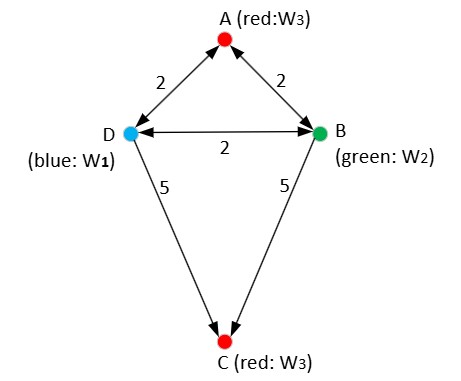}
    \label{fig:eg1_uncoded}
    }
    \subfigure[A Coded Scheme]{
    \includegraphics[width=0.4\linewidth]{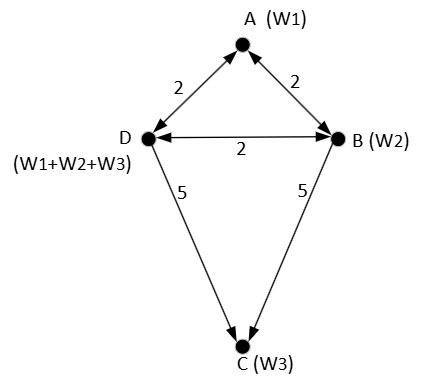}
    \label{fig:eg1_coded}
    }
    \caption{\small Uncoded and Coded Schemes for Example 1}
\end{figure}

We now turn our attention to average latency optimality. In ~\cite{acharya_etal_LatencyOptimalUncodedStorage_ISIT2024}, it was shown that when the file demands are equally likely at each node, if an uncoded storage is admissible on $\nng$, then it is also average latency optimal. This is because, for any uncoded admissible scheme on $\nng$, the set of $k$ latencies at each node $v$ take the minimum possible values, \(\{0, \lambda_{v,1}, \dots,\lambda_{v,(k-1)}\}\). Further, as all the demand probabilities are equal $p_{v,j} = \frac{1}{kn}$, the average latency $L_{\text{avg}}$ given by \eqref{eq:avg_latency}is also minimum since
\[L_{\text{avg}} = \frac{1}{kn}\sum_{v \in \mathcal{V}}\sum_{j \in [k]}\lambda_{v,(j-1)} \text{.}\] 
However, for generic file demand probabilities, this is not the case as shown in the example next.
\begin{figure}
    \subfigure[Nodes,RTTs]{
    \includegraphics[width=0.25\linewidth]{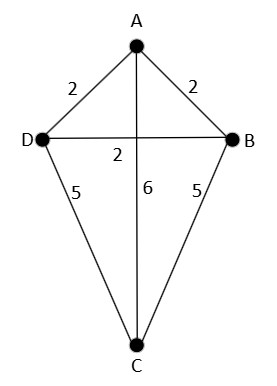}
    }
    \hfill
    \subfigure[$\nng$]{
    \includegraphics[width=0.25\linewidth]{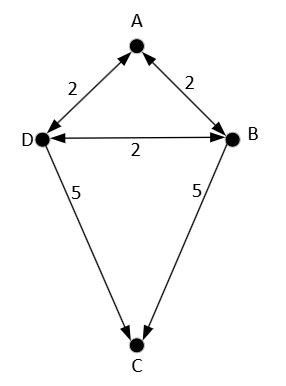}
    }
    \hfill
    \subfigure[Extended graph $\extg$]{
    \includegraphics[width=0.35\linewidth]{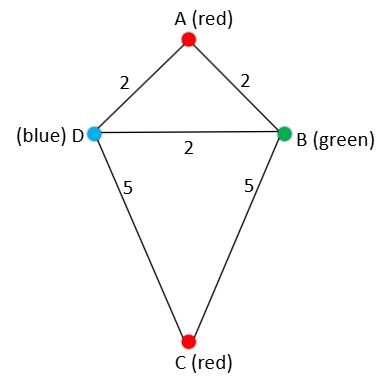}
    }
    \caption{\small Example~1: $(n,k) = (4,3)$ system.}
    \label{fig:eg1}
\end{figure}
\begin{example}
\label{example:n4k3}
Consider $(n,k)=(4,3)$ system as shown in Fig.~\ref{fig:eg1}, which also shows the nearest neighbor graph $\nng$ and the extended graph $\extg$. $\extg$ can be seen to be $(k=3)$-colorable and hence WC-latency optimal uncoded scheme exists. Consider the demand probability matrix as given in Table~\ref{tab:eg1_demand}. As a color is assoicated with a file, by exhaustively trying $3!=6$ ways of color-to-file assignment, we find the least average-latency uncoded storage as in Fig.~\ref{fig:eg1_uncoded}. Using the latency definitions given in Section~\ref{sec:sys_model}, the average latency turns out to be $L_{\text{avg}} =1.25$ units.
Now, consider a coded storage as in Fig.~\ref{fig:eg1_coded}. Note that, this is also admissible on $\nng$ and hence WC-latency optimal. Here, the average latency turns out to be $L_{\text{avg}} =0.95$, which is better than the best uncoded scheme. 
\end{example}
\begin{table}
\caption{\small Demand Probability $p_{v,j}$}
\begin{center}   
{\footnotesize
    \begin{tabular}{|c|c|c|c|}
    \hline
     Node $v$ & \multicolumn{3}{c|}{File $W_j$} \\
     & $W_1$ &$W_2$ &$W_3$   \\
     \hline
     A    & 0.2 & 0.025 & 0.025 \\
     B    & 0.025 & 0.2 & 0.025 \\
     C    & 0.025 & 0.025 & 0.2 \\
     D    & 0.025 & 0.025 & 0.2 \\
     \hline 
    \end{tabular}
    \label{tab:eg1_demand}
}
\end{center}

\end{table}
The reason for non-optimalilty of the uncoded scheme comes from the vertex coloring condition. In the example, the nodes $A$ and $C$ are forced to share the same file as they share the same color. However, $A$ and $C$ have different file preferences ($W_1$,$W_3$ respectively) causing latency to increase in one of the nodes. 
On the other hand, in a coded storage, there is a latency penalty at every coded node since there is no file that can be decoded with $0$-latency at this node. However, it does not have the vertex coloring restriction of the uncoded storage. Effectively, it turns out to be the better scheme.

Thus, we need to know additional conditions (apart from vertex-coloring on $\extg$) to be satisfied for an uncoded storage to be both WC-latency and average-latency optimal. While this is still open, we solve an interesting offshoot of the problem - Suppose we restrict to uncoded storage on $\nng$ (and hence WC-latency optimal); then what would be the optimal color to file assignment that results in least average latency? In the example above, the optimal assignment is as shown in Fig.~\ref{fig:eg1_uncoded}. Exhaustive search involves $k!$ possibilities. However, we come up with an efficient method where the problem is converted to that of well-known \emph{balanced assignment}, that takes polynomial time. This is explained in the next section.

\section{Optimal Color-to-File Assignment} \label{sec:optimal_file_to_color}

To find the average-latency optimal color-to-file assignment for uncoded schemes on $\nng$, we need to find the cost/contribution of a given color-to-file assignment towards the average-latency. Let $\phi:\mathcal{V} \to [k]$ be the node-to-file assignment corresponding to an admissible uncoded storage on $\nng$. The key idea here is that the average latency given by \eqref{eq:avg_latency} is the sum of probabilistically-weighted RTTs of all edges of $\nng$. So, one can obtain the sum by looking at either the incoming edges to the nodes, or the outgoing edges from the nodes.   

\begin{figure}
    \centering
	\subfigure[At a receive node $v$]{    
    \includegraphics[width=0.45\linewidth]{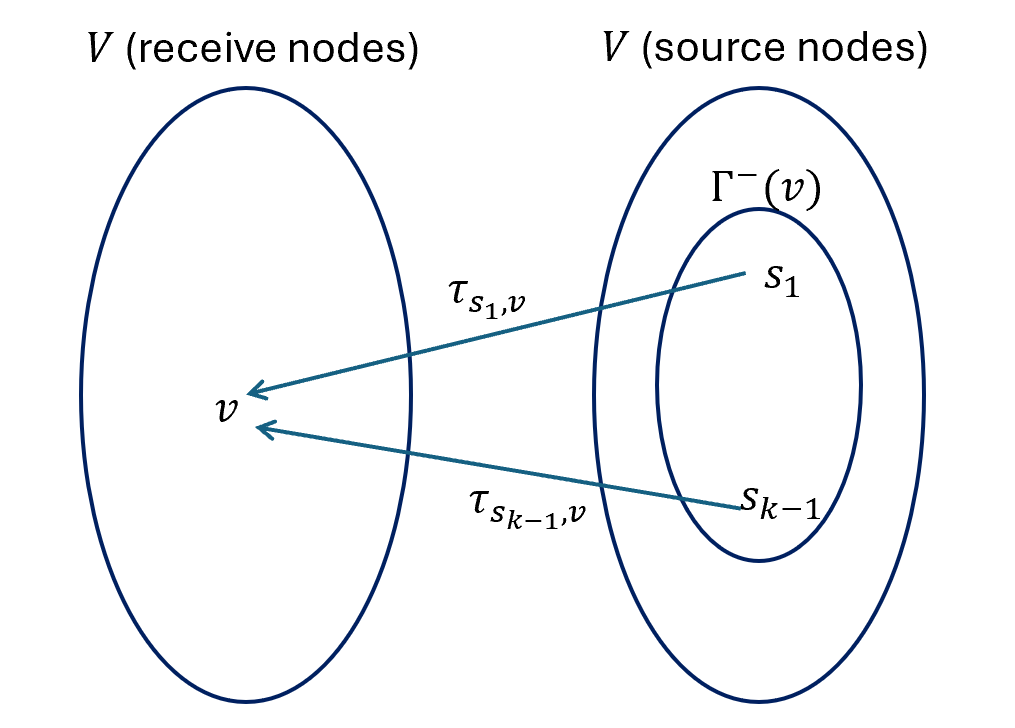}
    \label{fig:rx_node}
    }
	\subfigure[At a source node $s$]{    
    \includegraphics[width=0.45\linewidth]{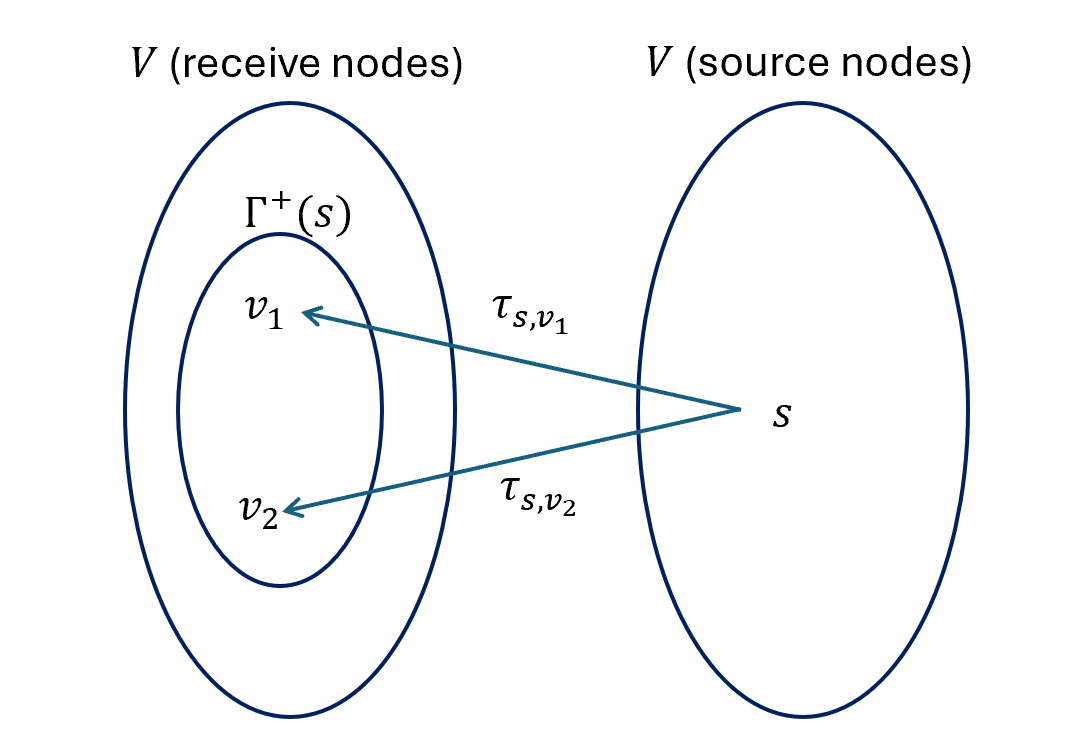}
    \label{fig:tx_node}
    }
    \caption{\small Bipartite equivalent of $\nng$ }
\end{figure}

We first adopt a receive-node perspective as shown in Fig.~\ref{fig:rx_node}. 
 Define \(\Gamma^-(v) = \{v\} \cup \{s \in \mathcal{V} , (s,v) \in \nng \}\), as the set of nodes with incoming edges to node $v$ (hence called source nodes). Note that the $k$ nodes in $\Gamma^{-}(v)$ together contain distinct $k$ files. Thus, with respect to a node $v$, there is one-one association between a file $W_j$ and the node $s \in \Gamma^{-}(v)$ satisfying $\phi(s)= j$. Also, since this is an uncoded scheme, each column of the recovery matrix $R^{(v)}(:,j)$ has a singe non-zero entry. Therefore, for this file-node combination $(W_j,s)$, we can see from \eqref{eq:latency_from_recovery_matrix} that $\ell_{v,j} = \tau_{s,v}$. Thus, the average latency becomes
\begin{equation}
    L_{\text{avg}} = \sum_{v\in \mathcal{V}}\sum_{j\in[k]} \ell_{v,j} p_{v,j} = \sum_{v\in \mathcal{V}} \sum_{s\in\Gamma^-(v)} \tau_{s,v}p_{v,\phi(s)}
    \label{eq:rx_latency}
    \text{.}
\end{equation}

We next look from the perspective of a source node $s$ as shown in Fig.~\ref{fig:tx_node}. Define \(\Gamma^+(s) = \{s\} \cup \{v \in \mathcal{V} , (s,v) \in \nng \}\), as the set of nodes with outgoing edges from node $s$. Define \emph{transmit latency} of node $s$ assigned with file $W_j$ as 
\begin{equation}
\ell_{\text{tx}}(s,j) := \sum_{v \in \Gamma^+(s)} \tau_{s,v} p_{v,j}\text{.}
\label{eq:tx_latency}
\end{equation}

The system-average latency is sum of the per-node transmit-latencies since
    {\small
    \begin{align}
        L_{\text{avg}} &= \sum_{v\in \mathcal{V}} \sum_{s\in\Gamma^-(v)} \tau_{s,v} p_{v,\phi(s)} \quad \text{(from \eqref{eq:rx_latency})} \nonumber \\
        & = \sum_{s\in \mathcal{V}}\sum_{v\in \Gamma^+(s)} \tau_{s,v} p_{v,\phi(s)} \text{  (as } s \in \Gamma^-(v) \iff v \in \Gamma^+(s)) \nonumber \\
	&= \sum_{s\in \mathcal{V}}\ell_{\text{tx}}(s,\phi(s))  \qquad \text{(from \eqref{eq:tx_latency})}\text{.}
	\label{eq:avg_from_tx_latency}
    \end{align}
    }

\begin{figure}
    \centering
    \includegraphics[width=0.6\linewidth]{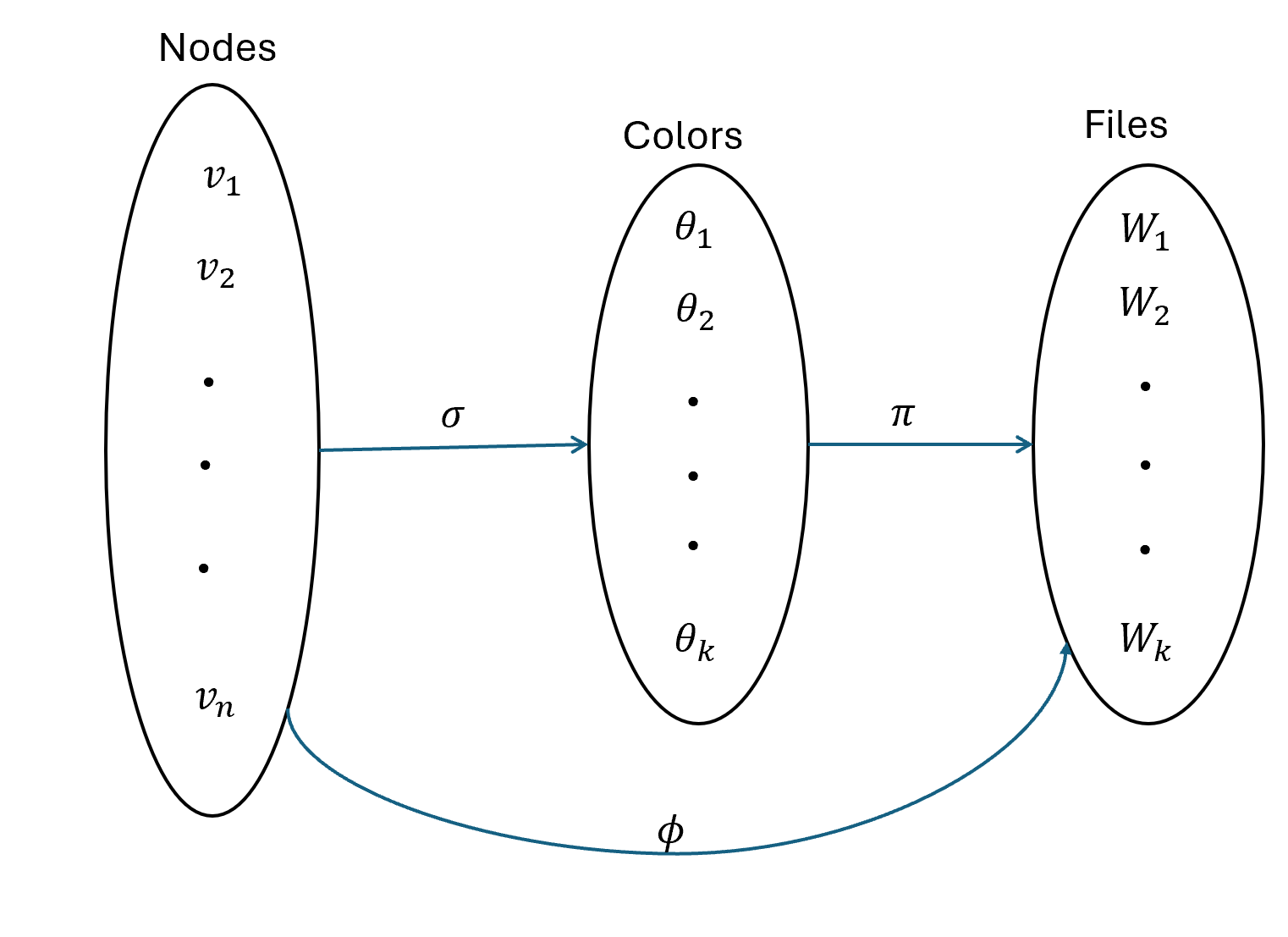}
    \caption{\small Node-to-Color-to-File Mapping}
    \label{fig:node_color_file_map}
\end{figure}

Note that the node-to-file mapping $\phi(.)$ is through the vertex coloring of $\extg$. Let \(\sigma: \mathcal{V} \to \Theta = \{\theta_1 \dots \theta_k\}\) be the vertex coloring map. Let \(\pi: \Theta \to [k]\) be the associated bijective mapping from colors to file indices. The node-to-file assignment is therefore \(\phi(v)  = \pi(\sigma(v))\) as shown in Fig.~\ref{fig:node_color_file_map}. Since a bunch of nodes with the same color are assigned the same file, we define the \emph{color-assignment cost} of color $\theta$ to file $W_j$, $c(\theta,j)$, as the sum of the transmit latencies of nodes assigned the same color $\theta$.
\begin{equation}
c(\theta,j) = \sum_{s\in \mathcal{V}, \sigma(s) = \theta}\ell_{\text{tx}}(s,j)\text{.}
\end{equation}

Now, the average latency expression in \eqref{eq:avg_from_tx_latency} can further be expressed in terms of color-assignment costs as follows.

\begin{align}
            L_{\text{avg}} & = \sum_{s\in \mathcal{V}}\ell_{\text{tx}}(s,\phi(s))
            = \sum_{s\in \mathcal{V}}\ell_{\text{tx}}(s,\pi(\sigma(s)))) \nonumber \\
            & = \sum_{\theta \in \Theta} \sum_{s\in \mathcal{V}, \sigma(s) = \theta}\ell_{\text{tx}}(s,\pi(\theta))  \quad \text{(where } \theta = \sigma(s)\text{)} \nonumber \\
            & = \sum_{\theta \in \Theta} c(\theta, \pi(\theta))\text{.}
            \label{eq:avg_latency_from_color_cost}
\end{align}

\begin{example*}
In Example~\ref{example:n4k3}, suppose we wish to find $\ell_{\text{tx}}(A,2)$, the transmit latency at node $A$ when assigned file $W_2$.  From $\nng$ in Fig.~\ref{fig:eg1}, we see that $A$ supplies files to the nodes $A$, $B$ and $D$, which are at RTTs $\{0,2,2\}$ respectively. Also, from Table~\ref{tab:eg1_demand}, the demand probabilities of file $W_2$ at the nodes $A,B,D$  are $\{0.025,0.2,0.025\}$ respectively. Hence \(\ell_{\text{tx}}(A,2) = \sum_{v\in \Gamma^+(A)}\tau_{A,v}p_{v,2} = 0(0.025)+2(0.2)+2(0.025) = 0.45\). In this manner, all entries of the $[\ell_{\text{tx}}(s,j)]$ matrix can be generated, as given in Table~\ref{tab:eg1_tx_latency}.
Also, from the vertex coloring of $\extg$ in Fig.~\ref{fig:eg1}, $A$ and $C$ are assigned same color (red). Hence the color-assignment cost for the color red is the sum \(c(\text{red},j) = \ell_{\text{tx}}(A,j)+ \ell_{\text{tx}}(C,j)\). The resulting color-cost matrix is given in Table~\ref{tab:eg1_cost_matrix}.   
\end{example*}

\begin{table}
\caption{\small Transmit latency matrix for Example~\ref{example:n4k3}}
 \begin{center}
   
 {\footnotesize
     \begin{tabular}{|c|c|c|c|}
          \hline
          Node & \multicolumn{3}{c|}{Transmit Latency} \\
          $s$ & $\ell_{\text{tx}}(s,1)$ & $\ell_{\text{tx}}(s,2)$ & $\ell_{\text{tx}}(s,3)$ \\
          \hline
          $A$ & $0.1$ & $0.45$ & $0.45$ \\
          $B$ & $0.575$ & $0.225$ & $1.45$ \\
          $C$ & $0$ & $0$ & $0$ \\
          $D$ & $0.575$ & $0.575$ & $1.1$ \\
          \hline
     \end{tabular}
 }
\end{center}

\label{tab:eg1_tx_latency}
 \end{table}

\begin{table}
\caption{\small Example~\ref{example:n4k3}: Color-Assignment Cost Matrix}
\begin{center}   
\footnotesize{
    \centering
    \begin{tabular}{|c|c|c|c|}
         \hline
         Color & \multicolumn{3}{c|}{Cost} \\
         $\theta$ & $c(\theta,W_1)$ & $c(\theta,W_2)$ & $c(\theta,W_3)$ \\
         \hline
         Red & $0.1$ & $0.45$ & $0.45$ \\
         Green & $0.575$ & $0.225$ & $1.45$ \\
         Blue & $0.575$ & $0.575$ & $1.1$ \\
         \hline
    \end{tabular}
}
\label{tab:eg1_cost_matrix}
\end{center}

\end{table}


Thus, from the expression in \eqref{eq:avg_latency_from_color_cost}, the uncoded storage problem with WC-latency optimality and average-latency optimal file assignment can be formulated as:
    \begin{enumerate}
        \item Find a valid $k$-vertex coloring on $\extg$: \(\sigma: \mathcal{V} \to \Theta\),
        \item And, find the bijective color-to-file map  \(\pi: \Theta \to [k]\) such that the average latency given below is minimized:
        \[
            L_{\text{avg}} =\sum_{\theta \in \Theta} c(\theta, \pi(\theta)) \text{.}
        \]
    \end{enumerate}

\begin{figure}
        \centering
        \subfigure[\small Complete Bipartite Graph]{
        \includegraphics[width=0.4\linewidth]{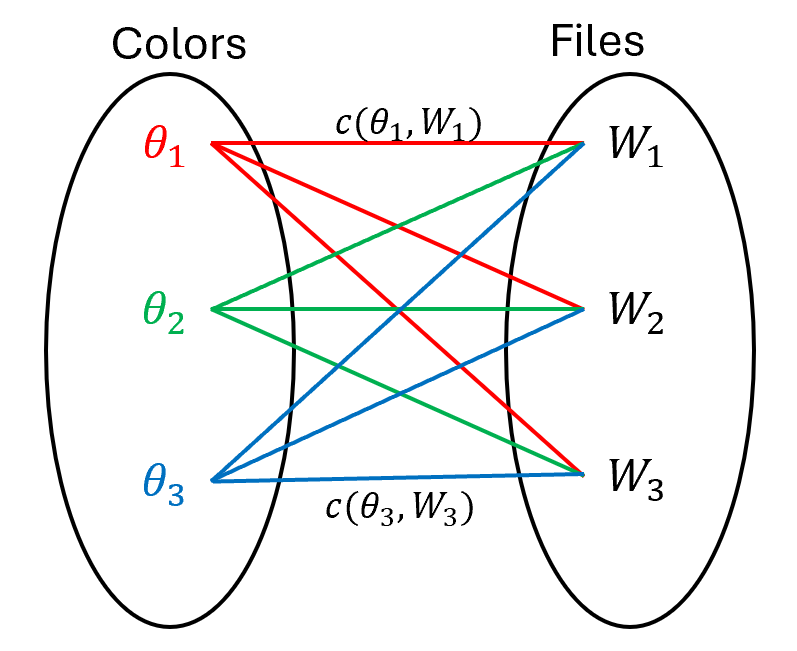}
\label{fig:bip_graph_main}
        }
        \subfigure[\small Maximal Matching]{
        \includegraphics[width=0.4\linewidth]{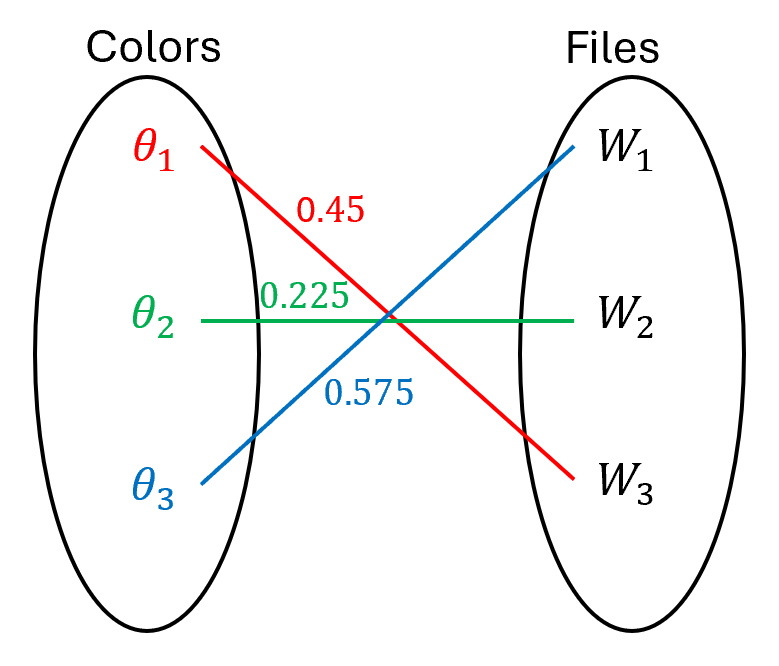}
\label{subfig:eg1_matching}
        }
        \caption{\small Balanced Assignment Formulation}

    \end{figure}

Finding the optimal bijective map $\pi(.)$ which minimizes the total cost is precisely the well-known \emph{balanced-assignment} problem\cite{hungarian_kuhn}. Here, the costs form the edge weights of a complete bipartite graph from the set of $k$ agents (colors) to the set of $k$ tasks (files) as in Fig.~\ref{fig:bip_graph_main}. Finding the optimal bijective map translates to finding a \emph{maximum matching} in the bipartite graph with minimum \emph{cost}\footnote{Maximum matching refers to  maximum number of independent edges. The cost of a matching is the sum of weights of the edges in the matching\cite{diestel}.}. There exist efficient algorithms (such as the \emph{Hungarian algorithm}\cite{hungarian_kuhn}\cite{hungarian_matrix_flood}) that solve this problem in polynomial time with  $O(k^3)$ complexity, as compared to $k!$ searches with brute-force method.
An application to Example~\ref{example:n4k3} is provided in the Appendix~\ref{app:balanced_assignment}, which gives the best assignment as in Fig.~\ref{subfig:eg1_matching}. The resulting average-latency is $L_{\text{avg}} = (0.45+0.225+0.575) = 1.25$, as asserted earlier in Section~\ref{sec:avg_latency}.

A final note concerns multiple vertex-coloring possibilities of $\extg$. In this case, we solve the balanced assignment problem for each choice of vertex-coloring, and then pick the choice that provides the least average-latency $L_{\text{avg}}$ given in \eqref{eq:avg_latency_from_color_cost}. 

\section{Conclusion}\label{sec:conc}

We considered the problem of finding uncoded storage schemes in a geo-distributed storage network with preferential file demands, that are optimal in terms of both WC-latency and average latency. WC-latency optimality is shown to be equivalent to a vertex coloring condition. With respect to average latency, identifying the conditions for existence of an uncoded scheme that performs better than any coded scheme is still open, except for the uniform demand case. Towards this, we have solved a relevant problem of finding the least average-latency scheme within the family of uncoded WC-latency optimal schemes, by modeling it as a balanced-assignment problem on a color-assignment cost matrix. Given that coded storage schemes can provide better average-latency than uncoded schemes in some networks, finding latency-optimal coded storage schemes is an interesting area for future work.

\newpage 
\bibliographystyle{IEEEtran}
\bibliography{references_GeoDistStorage}

\begin{thebibliography}{10}
\providecommand{\url}[1]{#1}
\csname url@samestyle\endcsname
\providecommand{\newblock}{\relax}
\providecommand{\bibinfo}[2]{#2}
\providecommand{\BIBentrySTDinterwordspacing}{\spaceskip=0pt\relax}
\providecommand{\BIBentryALTinterwordstretchfactor}{4}
\providecommand{\BIBentryALTinterwordspacing}{\spaceskip=\fontdimen2\font plus
\BIBentryALTinterwordstretchfactor\fontdimen3\font minus
  \fontdimen4\font\relax}
\providecommand{\BIBforeignlanguage}[2]{{%
\expandafter\ifx\csname l@#1\endcsname\relax
\typeout{** WARNING: IEEEtran.bst: No hyphenation pattern has been}%
\typeout{** loaded for the language `#1'. Using the pattern for}%
\typeout{** the default language instead.}%
\else
\language=\csname l@#1\endcsname
\fi
#2}}
\providecommand{\BIBdecl}{\relax}
\BIBdecl

\bibitem{Spanner}
J.~C. Corbett, J.~Dean, M.~Epstein, A.~Fikes, C.~Frost, J.~J. Furman,
  S.~Ghemawat, A.~Gubarev, C.~Heiser, P.~Hochschild, and W.~Hsieh, ``Spanner:
  Google’s globally distributed database,'' \emph{ACM Transactions on
  Computer Systems (TOCS)}, vol.~31, no.~3, p.~8, 2013.

\bibitem{aws-geo}
Https://aws.amazon.com/rds/aurora/global-database/.

\bibitem{cosmos-geo}
Https://learn.microsoft.com/en-us/azure/cosmos-db/distribute-data-globally.

\bibitem{dimakis2010network}
A.~G. Dimakis, P.~B. Godfrey, Y.~Wu, M.~J. Wainwright, and K.~Ramchandran,
  ``{Network Coding for Distributed Storage Systems},'' \emph{IEEE Transactions
  on Information Theory}, vol.~56, no.~9, pp. 4539--4551, 2010.

\bibitem{wu2009reducing}
Y.~Wu and A.~G. Dimakis, ``Reducing repair traffic for erasure coding-based
  storage via interference alignment,'' in \emph{2009 IEEE International
  Symposium on Information Theory}.\hskip 1em plus 0.5em minus 0.4em\relax
  IEEE, 2009, pp. 2276--2280.

\bibitem{gopalan2012locality}
P.~Gopalan, C.~Huang, H.~Simitci, and S.~Yekhanin, ``On the locality of
  codeword symbols,'' \emph{IEEE Transactions on Information theory}, vol.~58,
  no.~11, pp. 6925--6934, 2012.

\bibitem{tamo2014family}
I.~Tamo and A.~Barg, ``A family of optimal locally recoverable codes,''
  \emph{IEEE Transactions on Information Theory}, vol.~60, no.~8, pp.
  4661--4676, 2014.

\bibitem{rawat2016locality}
A.~S. Rawat, D.~S. Papailiopoulos, A.~G. Dimakis, and S.~Vishwanath, ``Locality
  and availability in distributed storage,'' \emph{IEEE Transactions on
  Information Theory}, vol.~62, no.~8, pp. 4481--4493, 2016.

\bibitem{balaji2018erasure}
S.~Balaji, M.~N. Krishnan, M.~Vajha, V.~Ramkumar, B.~Sasidharan, and P.~V.
  Kumar, ``Erasure coding for distributed storage: an overview,'' \emph{Science
  China Information Sciences}, vol.~61, no.~10, 2018.

\bibitem{vinayaketal2022FnT}
V.~Ramkumar, S.~Balaji, B.~Sasidharan, M.~Vajha, M.~N. Krishnan, and P.~V.
  vk~Kumar, ``Codes for distributed storage,'' \emph{Foundations and Trends in
  Communication and Information Theory}, vol.~19, no.~4, 2022.

\bibitem{cadambe}
V.~R. Cadambe and S.~Lyu, ``Brief announcement: {CausalEC}: {A} causally
  consistent data storage algorithm based on cross-object erasure coding,'' in
  \emph{2023 ACM Symposium on Principles of Distributed Computing}, 2023, pp.
  374--377.

\bibitem{cadambeArxiv}
------, ``Causal{EC}: {A} causally consistent data storage algorithm based on
  cross-object erasure coding,'' 2021, arXiv:2102.13310.

\bibitem{ardekani2014self}
\BIBentryALTinterwordspacing
M.~S. Ardekani and D.~B. Terry, ``{A Self-Configurable Geo-Replicated Cloud
  Storage System},'' in \emph{11th {USENIX} Symposium on Operating Systems
  Design and Implementation ({OSDI} 14)}.\hskip 1em plus 0.5em minus
  0.4em\relax Broomfield, CO: {USENIX} Association, Oct. 2014, pp. 367--381.
  [Online]. Available:
  \url{https://www.usenix.org/conference/osdi14/technical-sessions/presentation/ardekani}
\BIBentrySTDinterwordspacing

\bibitem{spanstore}
\BIBentryALTinterwordspacing
Z.~Wu, M.~Butkiewicz, D.~Perkins, E.~Katz-Bassett, and H.~V. Madhyastha,
  ``Spanstore: Cost-effective geo-replicated storage spanning multiple cloud
  services,'' in \emph{Proceedings of the Twenty-Fourth ACM Symposium on
  Operating Systems Principles}, ser. SOSP '13.\hskip 1em plus 0.5em minus
  0.4em\relax New York, NY, USA: Association for Computing Machinery, 2013, p.
  292–308. [Online]. Available: \url{https://doi.org/10.1145/2517349.2522730}
\BIBentrySTDinterwordspacing

\bibitem{replication-placement}
A.~{Jonathan}, M.~{Uluyol}, A.~{Chandra}, and J.~{Weissman}, ``Ensuring
  reliability in geo-distributed edge cloud,'' in \emph{2017 Resilience Week
  (RWS)}.\hskip 1em plus 0.5em minus 0.4em\relax Wilmington, DE, USA: IEEE,
  Sep. 2017, pp. 127--132.

\bibitem{shankaranarayanan2014performance}
P.~Shankaranarayanan, A.~Sivakumar, S.~Rao, and M.~Tawarmalani, ``Performance
  sensitive replication in geo-distributed cloud datastores,'' in \emph{2014
  44th Annual IEEE/IFIP International Conference on Dependable Systems and
  Networks}.\hskip 1em plus 0.5em minus 0.4em\relax Atlanta, GA, USA: IEEE,
  2014, pp. 240--251.

\bibitem{Abebe2018ECStoreBT}
\BIBentryALTinterwordspacing
M.~Abebe, K.~Daudjee, B.~Glasbergen, and Y.~Tian, ``Ec-store: Bridging the gap
  between storage and latency in distributed erasure coded systems,'' in
  \emph{2018 IEEE 38th International Conference on Distributed Computing
  Systems (ICDCS)}.\hskip 1em plus 0.5em minus 0.4em\relax Los Alamitos, CA,
  USA: IEEE Computer Society, jul 2018, pp. 255--266. [Online]. Available:
  \url{https://doi.ieeecomputersociety.org/10.1109/ICDCS.2018.00034}
\BIBentrySTDinterwordspacing

\bibitem{su2016systematic}
M.~Su, L.~Zhang, Y.~Wu, K.~Chen, and K.~Li, ``Systematic data placement
  optimization in multi-cloud storage for complex requirements,'' \emph{IEEE
  Transactions on Computers}, vol.~65, no.~6, pp. 1964--1977, 2016.

\bibitem{ec-geo-placement}
J.~{Matt}, P.~{Waibel}, and S.~{Schulte}, ``Cost- and latency-efficient
  redundant data storage in the cloud,'' in \emph{2017 IEEE 10th Conference on
  Service-Oriented Computing and Applications (SOCA)}.\hskip 1em plus 0.5em
  minus 0.4em\relax Kanazawa, Japan: IEEE, Nov 2017, pp. 164--172.

\bibitem{zare2022legostore}
H.~Zare, V.~R. Cadambe, B.~Urgaonkar, N.~Alfares, P.~Soni, C.~Sharma, and A.~A.
  Merchant, ``Legostore: a linearizable geo-distributed store combining
  replication and erasure coding,'' \emph{Proceedings of the VLDB Endowment},
  vol.~15, no.~10, pp. 2201--2215, 2022.

\bibitem{SoljaninEtAl}
G.~Joshi, Y.~Liu, and E.~Soljanin, ``On the delay-storage trade-off in content
  download from coded distributed storage systems,'' \emph{IEEE Journal on
  Selected Areas in Communications}, vol.~32, no.~5, pp. 989--997, 2014.

\bibitem{lee_shah_huang_ramachandran_2017}
K.~Lee, N.~B. Shah, L.~Huang, and K.~Ramchandran, ``The mds queue: Analysing
  the latency performance of erasure codes,'' \emph{IEEE Transactions on
  Information Theory}, vol.~63, no.~5, pp. 2822--2842, 2017.

\bibitem{acharya_etal_LatencyOptimalUncodedStorage_ISIT2024}
S.~Acharya, P.~V. Kumar, and V.~R. Cadambe, ``On existence of latency optimal
  uncoded storage schemes in geo-distributed data storage systems,'' in
  \emph{2024 IEEE International Symposium on Information Theory (ISIT)}, 2024,
  pp. 1462--1467.

\bibitem{diestel}
R.~Diestel, \emph{{Graph Theory}}, 5th~ed.\hskip 1em plus 0.5em minus
  0.4em\relax Springer books, 2004.

\bibitem{hungarian_kuhn}
\BIBentryALTinterwordspacing
H.~W. Kuhn, ``The hungarian method for the assignment problem,'' \emph{Naval
  Research Logistics Quarterly}, vol.~2, no. 1-2, pp. 83--97, 1955. [Online].
  Available:
  \url{https://onlinelibrary.wiley.com/doi/abs/10.1002/nav.3800020109}
\BIBentrySTDinterwordspacing

\bibitem{hungarian_matrix_flood}
\BIBentryALTinterwordspacing
M.~M. Flood, ``The traveling-salesman problem,'' \emph{Operations Research},
  vol.~4, no.~1, pp. 61--75, 1956. [Online]. Available:
  \url{https://EconPapers.repec.org/RePEc:inm:oropre:v:4:y:1956:i:1:p:61-75}
\BIBentrySTDinterwordspacing

\bibitem{goemans_mit_lec_notes}
M.~X. Goemans, ``Lecture notes on bipartite matching,'' February 2013, 18.433
  Combinatorial Optimization, Massachusetts Institute of Technology.

\end{thebibliography}

\newpage
\appendices 
\section{Balanced Assignment Problem} \label{app:balanced_assignment}
We apply the matrix version of the Hungarian algorithm given in \cite{hungarian_matrix_flood} to Example~\ref{example:n4k3} of Section~\ref{sec:avg_latency}, whose color-assignment cost matrix is as given in Table~\ref{tab:eg1_cost_matrix}. 
In a general setting, given $k$ agents (colors in this case) denoted by the set $A= \{\theta_1 \dots\theta_k\}$, $k$ tasks (files in this case) given by $B = \{W_1 \dots W_k\}$, and the cost matrix $C = [c_{i,j}]_{i,j \in [k]}$ providing the cost associated with every agent-task combination, the objective is to:
\bit
\item [] Find bijection \(\pi: A \to B \) that minimizes the following:
    \[f(\pi) = \sum_{\theta \in A} c(\theta,\pi(\theta)) \text{.}\] 
\eit

\begin{figure}[H]
        \centering
        \subfigure[\small Complete Bipartite Graph]{
        \includegraphics[width=0.4\linewidth]{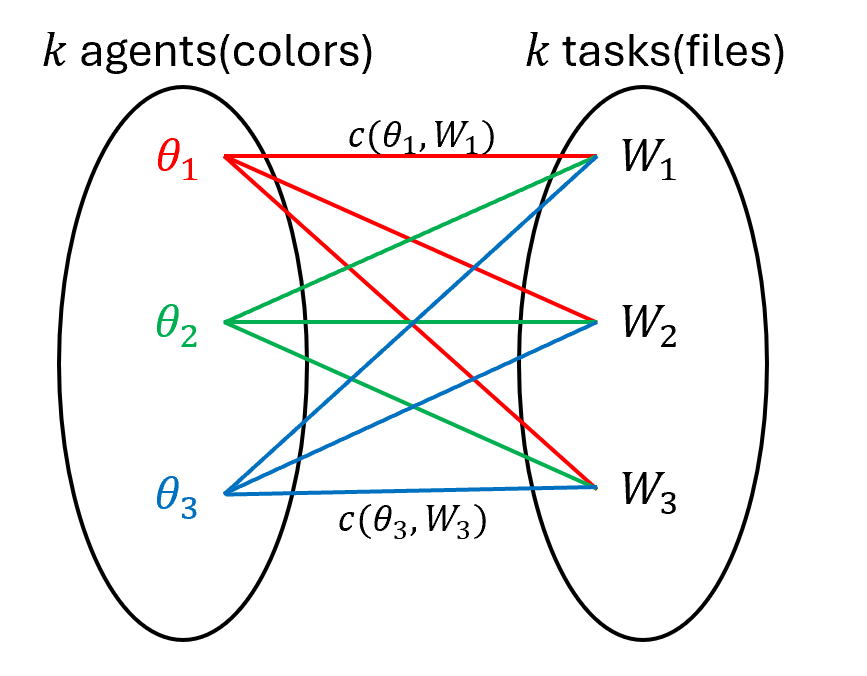}
        \label{subfig: bip_graph}
        }
        \subfigure[\small A Maximum Matching of size $k=3$]{
        \includegraphics[width=0.4\linewidth]{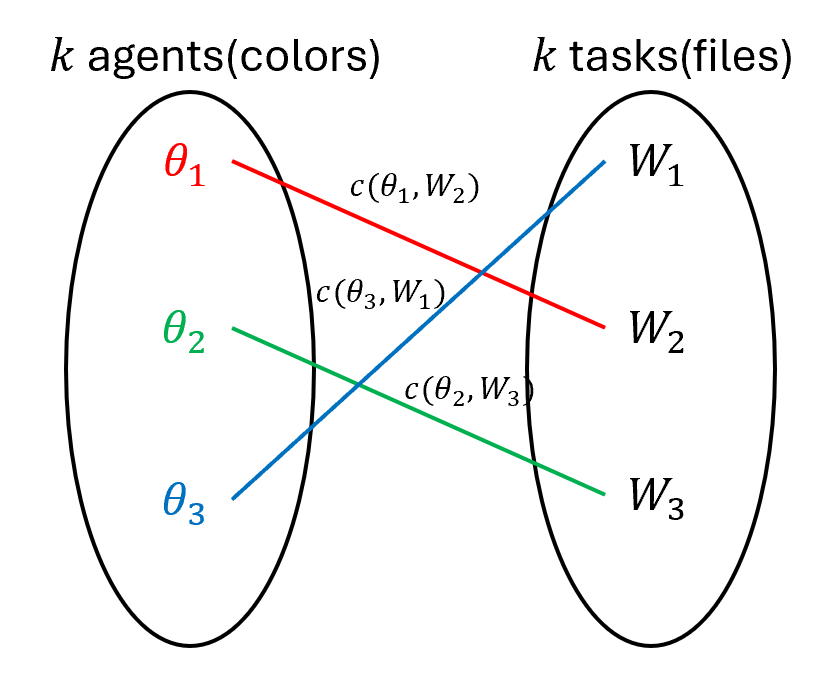}
        \label{subfig:eg_max_matching}
        }
        \caption{\small Balanced Assignment Formulation}
    \end{figure}

Consider a complete  weighted bipartite graph $\mathcal{G}$ on the vertex set $\{A, B\}$, with the edge $(i,j), i \in A, j \in B$ assigned a weight $c_{i,j}$, as shown in Fig.~\ref{subfig: bip_graph}. A matching in this graph is defined as a set of independent edges (edges with no endpoints in common). Maximum (cardinality) matching refers to a matching with maximum size ($k$, in this case). A possible maximum matching for the current example is given in Fig.~\ref{subfig:eg_max_matching}. Now, we observe that a bijection \(\pi: A \to B \) is equivalent to a maximum matching in the bipartite graph $\mathcal{G}$. The objective is therefore to find a maximum matching with minimum cost. Here, the cost of a matching is defined as the sum of the weights of the edges in the matching. Since the sets $A$ and $B$ have the same size, finding the optimal assignment (or matching) $\pi(.)$ is known as the \emph{balanced} assignment problem.

The main observation employed by the algorithm is that optimal matching does not change if a constant $\Delta$ is removed from any row/column of the cost matrix. The reason is that the total cost changes from \(f(\pi) \to f(\pi) - \Delta\) for all assignments $\pi(.)$, and therefore does not alter the optimal $\pi(.)$.  Thus, the algorithm proceeds as follows, along with the application to Example~\ref{example:n4k3}.

\subsection*{
\textbf{Step~$1$: Get a non-negative entry cost-matrix $C$.}
\bit
\item Subtract each row with its smallest value.
\eit
}
\[
       \begin{bmatrix}
           \circled{0.1} & 0.45 & 0.45 \\
          0.575 & \circled{0.225} & 1.45 \\
          \circled{0.575} & 0.575 & 1.1 
       \end{bmatrix} 
       \to 
        \begin{bmatrix}
           0 & 0.35 & 0.35 \\
          0.35 & 0 & 1.225 \\
          0 & 0 & 0.525 
       \end{bmatrix} 
    \]
    
    The smallest entry in each row is circled as shown. With this step, observe that all entries are $0$ or positive. The equivalent bipartite graph $\mathcal{G}$ has non-negative edge weights. Now, form a bipartite subgraph on the same node sets $\{A,B\}$, having only $0$-cost edges (which we term as \emph{tight} edges). We term this as \emph{cost-tight subgraph} $\mathcal{T}$ since the cost of any of its edges is the minimum value of $0$. In this graph, suppose that there exists a matching of size $k$. Then, this is an optimal matching since its cost is $0$ which is the minimum (as the cost of any other matching is $\ge 0$).
    
    Thus, the modified equivalent objective is to find a matching of size $k$ in the cost-tight subgraph $\mathcal{T}$. 
        
\subsection*{
        {\textbf{Step~$2$: Find maximum matching $M$ in the cost-tight subgraph $\mathcal{T}$.}}
    \begin{itemize}
        \item If $M$ is of size $k$, terminate the algorithm: 
        \bit
            \item The matching $M$ in $\mathcal{T}$ is also a maximum matching in $\mathcal{G}$ since its size is $k$. 
            \item Further, as the matching $M$ is made of $0$-cost edges, it has minimum cost.
        \eit
        \item Else, go to Step~$3$.
    \end{itemize}
}
\begin{figure}
    \centering
    \subfigure[\small Cost-Tight Subgraph $\mathcal{T}$ with $0$-cost edges.]{
    \includegraphics[width=0.4\linewidth]{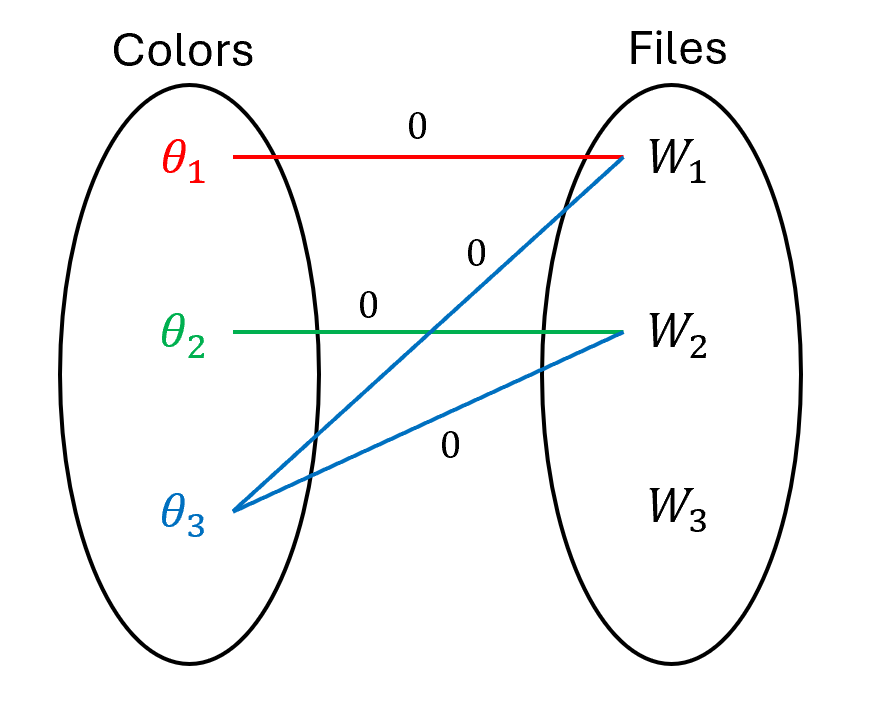}
    \label{subfig:step2_cost_tight_subgraph}
    }
    \hfill
    \subfigure[\small Maximum matching in $\mathcal{T}$.]{
    \includegraphics[width=0.4\linewidth]{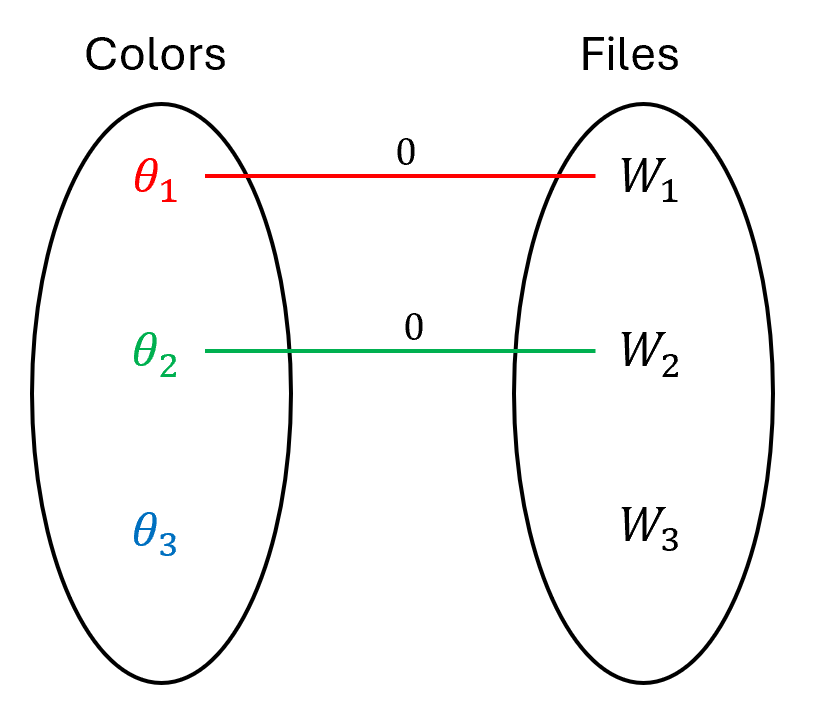}
    \label{subfig:step2_maximum_matching}
    }
    \caption{\small Step~$2$: matching in the cost-tight subgraph.}
    \end{figure}

    For the current example, the cost-tight subgraph is as shown in Fig.~\ref{subfig:step2_cost_tight_subgraph}. Note that a matching always exists in this graph since the number of tight edges is non-zero (at least $k$, one on each row). In order to find a matching of maximum size, standard techniques such as Kuhn's algorithm~\cite{hungarian_kuhn} can be employed. The maximum matching in the cost-tight subgraph is as shown in Fig.~\ref{subfig:step2_maximum_matching} which has size $r=2$, strictly less than $k=3$. Hence we proceed to the next step. 
        
    \noindent \textbf{Note:} By subtracting the last column with its smallest element, we could have obtained another zero-valued entry that would result in matching of $k=3$. However, for the sake of explaining the algorithm for a more generic case where this is not possible, we omit this observation and show further steps.

    \subsection*{\textbf{Step~$3$: Find a minimum vertex cover on the cost-tight subgraph $\mathcal{T}$.}}
    
    In order to increase the size of matching in cost-tight subgraph, we need to create additional tight edges (that is, $0$-cost entries) without affecting the existing matching. For this, we first find a minimum vertex cover on the cost-tight subgraph. 
    \begin{defn}
    A vertex cover of a graph is a set of vertices that contain at least one endpoint of all edges in the graph. A Minimum vertex cover is a vertex cover with the minimum size.  
    \end{defn} 
    In the current context, the vertex cover of the cost-tight subgraph should contain one end point of all the $4$ tight edges of Fig.~\ref{subfig:step2_cost_tight_subgraph}. Also a (minimum) vertex cover should contain at least one endpoint of the matched edges, and hence has size at least as much as the size of (maximum) matching. However, K\"onig's theorem\cite{diestel} goes further and asserts that the maximum cardinality of a matching in a bipartite graph is equal to the minimum cardinality of a vertex cover of its edges \footnote{
    \begin{itemize}
        \item Section~\ref{subsec:vertex cover} provides a proof of the theorem by explicitly constructing a vertex cover whose size is that of the maximum matching. 
        \item Any minimum vertex cover of the cost-tight subgraph $\mathcal{T}$ can be used for the algorithm to work. However, by choosing a specific vertex cover detailed in Section~\ref{subsec:vertex cover}, polynomial time convergence can be obtained. The convergence is discussed in Section~\ref{subsec:hungarian_convergence}.
    \end{itemize}
    }.
    Coming back to our example, K\"onig's Theorem indicates that there exists a minimum vertex cover of size $2$ in the cost-tight subgraph $\mathcal{T}$ of Fig.~\ref{subfig:step2_cost_tight_subgraph}. We can see that $V_c =\{W_1,W_2\}$ is a minimum vertex cover of $\mathcal{T}$. 
      
   We now observe the following equivalence between the entries in matrix $C$ and the bipartite graph $\mathcal{T}$.
   \begin{itemize}
       \item A $0$-valued entry in $C$ corresponds to a tight-edge in $\mathcal{T}$.
       \item A row $i$ in $C$ is equivalent to a vertex $i$ in the set $A$.
       \item A column $j$ in $C$ is equivalent to a vertex $j$ in the set $B$.
       \item Thus, a minimum vertex cover $V_c$ in $\mathcal{T}$ is equivalent to covering all $0$-valued entries  in the cost matrix $C$ with minimum number of horizontal lines (rows) and vertical lines (columns).
   \end{itemize}
   
   We therefore \textbf{mark} all the rows/columns corresponding to the minimum vertex cover $V_c$. In the example, columns $\{1,2\}$  corresponding to $V_c =\{W_1,W_2\}$ cover all $0$s of the cost-matrix $C$ as shown below:
 \[\begin{bmatrix}             \color{orange}{\underline{\text{col1}}} & \color{magenta}{\underline{\text{col2}}} & \\                      \color{orange}{0} & \color{magenta}{0.35} & 0.35 \\             
\color{orange}{0.35} & \color{magenta}{0} & 1.225 \\             
\color{orange}{0} & \color{magenta}0 & 0.525          \end{bmatrix} \text{.}        \]

\subsection*{
\textbf{Step~$4$: Create additional tight-edges as follows:} 
\begin{enumerate}
    \item Find the smallest \emph{unmarked} element $\Delta$ where unmarked elements are those which are not included by vertex-cover lines.
    \item Subtract $\Delta$ from all \textbf{unmarked} rows. 
    \item Add $\Delta$ to all \textbf{marked} columns.
\end{enumerate} 
}
 
The smallest unmarked element is $\Delta = 0.35$ in the example (circled in the cost-matrix below). Therefore, subtract this from all unmarked rows (all $3$ rows in the example). This results in negative numbers in the marked columns $1$, $2$. However, by adding $\Delta$ back to the marked columns, all entries have again become non-negative as shown below. Note that the unmarked entries are not distubed in the last step. Thus, with Step~$4$, the entry corresponding to the smallest unmarked element $\Delta$ has become $0$, and hence we have a new tight-edge.
        \[
        \begin{bmatrix}
           0 & 0.35 & \circled{0.35} \\
          0.35 & 0 & 1.225 \\
          0 & 0 & 0.525 
       \end{bmatrix} 
       \to 
        \begin{bmatrix}
           -0.35 & 0 & \circled{\color{red}{0}} \\
          0 & -0.35 & 0.875 \\
          -0.35 & -0.35  & 0.175 
       \end{bmatrix}  
        \to
        \begin{bmatrix}
           \color{red}{0} & 0.35 & \circled{\color{red}{0}} \\
          0.35 & \color{red}{0} & 0.875 \\
          \color{red}{0} & \color{red}{0} & 0.175 
       \end{bmatrix}\text{.}
     \]

\subsection*{
    \textbf{Step~$5$ (same as Step~$2$): Find maximum matching $M$ in the updated cost-tight subgraph $\mathcal{T}$.}
    \begin{itemize}
        \item If $M$ is of size $k$, terminate the algorithm. 
        \bit
            \item The matching $M$ in $\mathcal{T}$ is also a maximum matching in $\mathcal{G}$ since its size is $k$. 
            \item Further, as the matching $M$ is made of $0$-cost edges, it has minimum cost.
        \eit
        \item Else, repeat Steps~$\{3,4,5\}$ until we get a maximum matching of size $k$.
    \end{itemize}
}
 
        \begin{figure}
        \vspace{-1em}
        \hfill
        \subfigure[\small Cost-Tight Subgraph $\mathcal{T}$ of Step~$4$]{
        \includegraphics[width=0.36\linewidth]{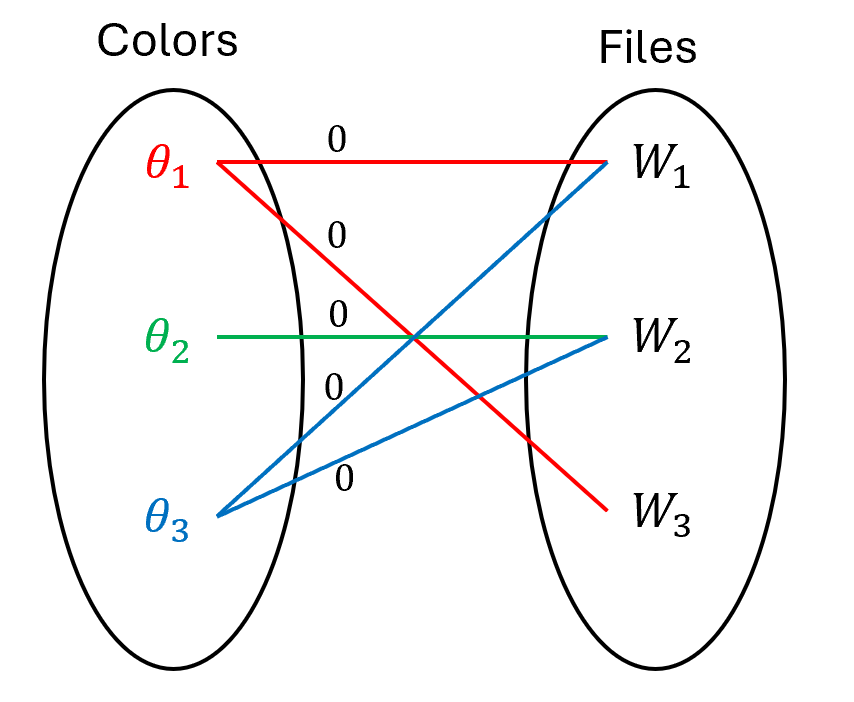}
        }
        \hfill
        \subfigure[\small Maximum matching in $\mathcal{T}$]{
        \includegraphics[width=0.36\linewidth]{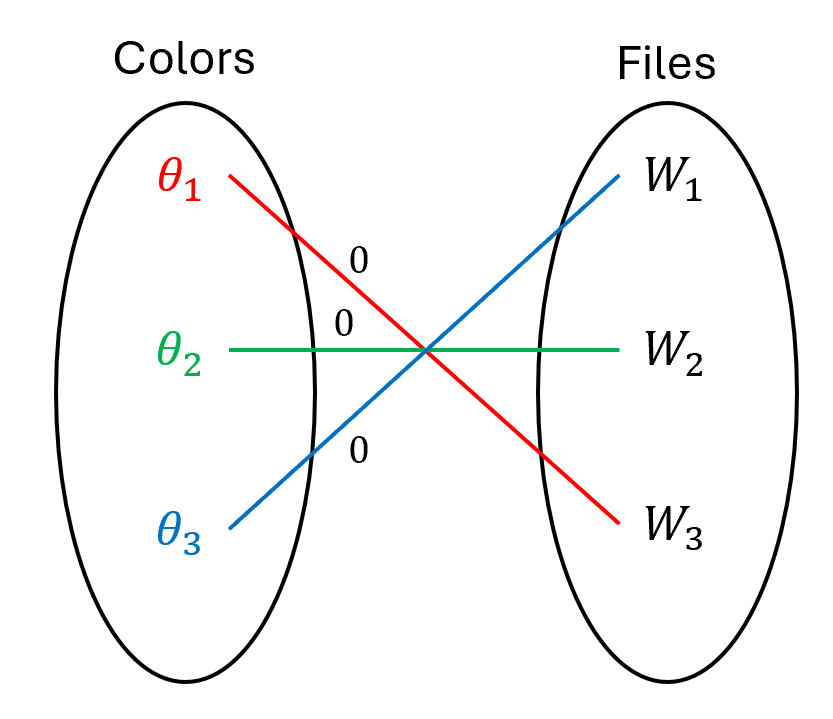}
        }
        \caption{\small Step~$5$: matching in the cost-tight subgraph $\mathcal{T}$.}
        \label{fig:0-edge-matching-final}
    \end{figure}

For the example, the updated cost-tight subgraph $\mathcal{T}$ and its maximum matching $M$ are as shown as shown in Fig.~\ref{fig:0-edge-matching-final}. Note that $(\theta_1, W_3)$ is the newly added tight-edge. The matching is of size $k=3$ and has minimum cost of $0$, and hence is optimal.
Thus, the optimal assignment corresponding to the maximum matching is \[\pi(\theta_1) = W_3,\pi(\theta_2) = W_2,\pi(\theta_3) = W_1 \text{.}\]

    \begin{table}
\footnotesize{
    \centering
    \begin{tabular}{|c|c|c|c|}
         \hline
         Color & \multicolumn{3}{c|}{Cost} \\
         $\theta$ & $c(\theta,W_1)$ & $c(\theta,W_2)$ & $c(\theta,W_3)$ \\
         \hline
         Red & $0.1$ & $0.45$ & \circled{$0.45$} \\
         Green & $0.575$ & \circled{$0.225$} & $1.45$ \\
         Blue & \circled{$0.575$} & $0.575$ & $1.1$ \\
         \hline
    \end{tabular}
    \caption{\small Optimal assignment Color-Cost Matrix}
    \label{tab:optimal_cost}
}
\end{table}

\noindent The resulting optimal cost of the assignment can be seen from Table~\ref{tab:optimal_cost} as \[L_{\text{avg}} = \sum_{\theta \in A} c(\theta, \pi(\theta)) = 0.45+0.225+0.575 = 1.25 \text{.}\] 

\subsection{\textbf{Convergence of the Algorithm}}
\label{subsec:hungarian_convergence}

We next show that the above algorithm terminates in finite number of iterations. 
\begin{itemize}
    \item In Step~$3$, let $r$ be the size of the maximum matching $M$ of the cost-tight subgraph $\mathcal{T}$. We consider $r \le (k-1)$ case (since the algorithm terminates when $r=k$).
    \item The number of marked lines (rows or columns) is equal to the size of the minimum vertex cover, which in turn is equal to the size of the maximum matching, $r$, as per K\"onig's theorem. 
    \item Now, in Step-$4$, a positive value of $\Delta$ is subtracted from all unmarked rows, and then added to all marked columns. 
    \item This is equivalent to subtracting $\Delta$ from all rows (including marked rows), and then adding $\Delta$ to all marked rows and columns.
    \item Now, define total cost of the cost-matrix $C$ as the sum of all its entries, $S = \sum_{i,j \in [k]} c_{i,j}$. Thus, in Step~$4$, since $\Delta >0 $ is subtracted from $k$ rows and added to $r$ marked rows/columns,  the total cost changes to
    \[
    S_{\text{new}} = S_{\text{old}} -\Delta(k^2)+\Delta(kr) = S_{\text{old}} -k \Delta(k-r) < S_{\text{old}}\text{.} 
    \]
    \item Since the total cost $S$ strictly reduces and is non-negative, the algorithm terminates. Also, note that $\Delta$ is strictly bounded below by the least difference between two entries in the original cost matrix. Hence, the algorithm terminates in finite iterations. 
\end{itemize}
The above argument does not show the polynomial running time of the algorithm. To show this, we choose a specific minimum vertex cover in Step~$3$, along with an incremental matching procedure in Step~$5$, as described next.

\subsection{\textbf{Finding minimum vertex cover given a maximum matching}}
\label{subsec:vertex cover}

    Here, we provide the construction of a minimum vertex cover on a given bipartite graph with maximum matching. In the process, we also prove the K\"onig's theorem stated below.
        \begin{theorem} [K\"onig\cite{diestel}]
        The maximum cardinality of a matching in a bipartite graph is equal to the minimum cardinality of a vertex cover of its edges. 
    \end{theorem}
    \begin{proof}    

    Let $M \subseteq E$ be a matching in the bipartite graph  $(\{A,B\},E)$. To obtain the minimum vertex cover, we define the following\cite{diestel}. 
    \begin{enumerate}
        \item Unmatched vertices: The vertices in $\{A,B\}$ that are not endpoints of any of the matched edges in $M$.
        \item Unmatched edges: The edges in $E$ not in the matching $M$.
        \item Alternating Path: This is a directed path starting from an unmatched vertex in $A$ alternating between unmatched edge from $A$ to $B$ and a matched edge from $B$ to $A$. 
      \item Augmenting Path:  If an alternating path (starting from an unmatched vertex in $A$) ends in an unmatched vertex in $B$, it is called as an augmenting path. Note that augmenting path has odd number of edges, with number of unmatched edges one more than the number of matching edges.  
    \end{enumerate}
       Now, let $M$  be a \textbf{maximum} matching. With this $M$, there cannot be any augmenting path. This is because, by swapping unmatched edges and the matched edges of the augmenting path, we can increase the number of matching edges by $1$, contradicting the fact that the matching is of maximum cardinality. 

    Let $L$ be the set of vertices that are part of some alternating path starting from any unmatched vertex in $A$. Consider the set of vertices given by $V_c = (A-L) \cup (B\cap L)$. This is a vertex cover from the following observation.
      \begin{itemize}
          \item Since the bipartite graph has edges between the sets $A$ and $B$, any edge is incident to a vertex in set $A$. 
          \item Now, $A = (A-L) \cup (A\cap L)$.
          \item If an edge is incident to $(A-L)$, then it is covered by $V_c$.
          \item On the other hand, if the edge is incident to $(A\cap L)$, then it joins a matched vertex in $B$. (If not, this edge connects to an unmatched vertex in $B$, resulting in an augmenting path. This contradicts the fact that $M$ is a maximum matching.) Hence, the edge is part of an alternating path. Therefore, by definition of $L$, the other endpoint of the edge belongs to $B\cap L$, and hence in $V_c$. 
      \end{itemize}
      Note that, by construction, the vertex cover $V_c$ contains exactly one end point of each of the matched edges - the endpoint in $B$ if the matched edge belonging to $L$, else the endpoint from $A$. Hence the $\lvert V_c \rvert = \lvert M\rvert $. Since any vertex cover should contain at least one end-point of matching edges (and hence has size $\ge \lvert M\rvert$), $V_c$ is indeed the minimum vertex cover.
      
      The above construction thus also proves the K\"onig's theorm. 
      \end{proof}

\subsection{\textbf{Improvement to Polynomial Running Time}}
Here, we describe certain modifications to the algorithm described earlier, with the objective for obtaining a polynomial running time. For further details, please refer to \cite{goemans_mit_lec_notes} and references therein. 
\ \\
\subsubsection*{(Improved) Step~$3$ - Choice of Minimum Vertex Cover $V_c$}
\ \\
Here, we choose the minimum vertex cover $V_c$ obtained exactly using the procedure in Section~\ref{subsec:vertex cover}. By choosing this vertex cover, the maximum matching $M$ obtained in an iteration will continue to be a valid matching (not necessarily maximum) for the next iteration for the following reason:
\begin{itemize}
    \item In Step-$4$, it can be seen that the entries that are modified in the marked rows/columns are those lying in the intersection of marked rows and marked columns. These entries increase by $\Delta$ whereas rest of the entries are unchanged. 

    \item By construction, the vertex $V_c$ contains precisely one of the endpoints from each edge of the maximum matching $M$. 
    \item Equivalently, in the cost-matrix $C$, the the $0$-cost entries corresponding to the matched edges do not fall in the intersection of marked rows and marked columns. Hence these entries remain $0$ in the updated cost-matrix. 
    \item Hence, the edges of the previous matching $M$ remain as tight-edges in the updated cost-tight subgraph.
\end{itemize}
\ \\
\subsubsection*{(Improved) Step~$5$ - Maximum Matching $M$}
\ \\
In this step, instead of freshly finding a maximum matching in the updated cost-tight subgraph $\mathcal{T}$, we continue to build upon the matching $M$ of the previous step, as it is still a valid matching as explained earlier. But the cost-tight subgraph $\mathcal{T}$ now has at least one additional tight-edge, say $(i,j)$, obtained from Step~$4$. Thus there are $2$ possibilities:
\begin{enumerate}
    \item The additional tight-edge creates an augmenting path from an unmatched vertex in $A$ (with the existing matching $M$). In this case, we simply obtain a new matching $M'$ with size one more than $M$ by swapping the unmatched and matched edges of the augmenting path.
    \item There exists no augmenting path from an unmatched vertex in $A$. Hence, the previous matching $M$ remains a maximum matching. But, the vertex $j$ of the new tight edge $(i,j)$ can be reached via an unmatched vertex in $A$. By the vertex cover construction described in Section~\ref{subsec:vertex cover}, the vertex $j$ is now part of the new vertex cover $V_c$. Equivalently, in the cost-matrix $C$, the number of marked columns has increased by $1$.
\end{enumerate}
\ \\
\subsubsection*{Complexity}
\begin{itemize}
    \item As explained above, at each iteration, either the size of matching increases,  or the number of marked columns increases. Since there are only $k$ columns, the size of the matching increases (at least by $1$) certainly in $k$ iterations. Therefore, the algorithm terminates in $O(k^2)$ iterations when matching of size $k$ is obtained.
    \item Using depth-first search method (see \cite{goemans_mit_lec_notes}), it can be shown  that the process of finding an augmenting path (or the absence of it) on an existing matching $M$ can be done in $O(|E|)$ steps where $E$ is the edge set of $\mathcal{T}$ . Thus, each iteration takes $O(k^2)$ steps.
    \item Combining the two observations above, the entire algorithm converges in $O(k^4)$ running time. By looking more closely into how the maximum matching and minimum vertex cover changes between the two successive iterations\cite{goemans_mit_lec_notes}, one can reduce the running time to $O(k^3)$. 
    
\end{itemize}

\end{document}